\documentclass[11pt,a4paper]{article}

\pdfoutput=1

\usepackage{jheppub} 
\usepackage{xcolor}
\usepackage[T1]{fontenc}
\usepackage{tocloft}
\usepackage[normalem]{ulem}
\usepackage{subcaption}
\usepackage{array}
\usepackage{enumitem}
\usepackage{extarrows}
\usepackage{mathtools}
\usepackage{slashed} 
\usepackage{physics} 

\allowdisplaybreaks[1] 

\newcommand{\eps}{\epsilon}

\newcommand{\no}{\nonumber}
\newcommand{\ptv}{p_T^{\rm veto}}

\newcommand{\ns}{{\slashed{n}}}
\newcommand{\nsb}{{\slashed{\bar{n}}}}

\newcommand{\als}{\alpha_s}

\title{\boldmath 
The NNLO quark beam function for jet-veto resummation}

\author{Guido Bell, Kevin Brune, Goutam Das, and Marcel Wald}

\note{SI-HEP-2022-13, P3H-22-064.}

\affiliation{Theoretische Physik 1, 
Center for Particle Physics Siegen, 
Universit{\"a}t Siegen,\\
Walter-Flex-Stra{\ss}e 3, 57068 Siegen, Germany}
\emailAdd{bell@physik.uni-siegen.de}
\emailAdd{brune@physik.uni-siegen.de}
\emailAdd{goutam.das@uni-siegen.de}
\emailAdd{marcel.wald@uni-siegen.de}

\abstract{We consider the quark beam function that describes collinear initial-state radiation that is constrained by a veto on reconstructed jets. As the veto is imposed on the transverse momenta of the jets, the beam function is subject to rapidity divergences, and we use the collinear-anomaly framework to extract the perturbative matching kernels to next-to-next-to-leading order (NNLO) in the strong-coupling expansion. Our calculation is based on a novel framework that automates the computation of beam functions in Mellin space and it provides the ingredients to extend jet-veto resummations for quark-initiated processes to NNLL$'$ accuracy.
}

\keywords{QCD, Soft-Collinear Effective Theory, NNLO Computations}
\arxivnumber{2207.05578}

\begin{document} 

\maketitle
 
\flushbottom

%%%%%%%%%%%%%%%%%%%%%%%%%%%%%%%%%%%%%%%%%%%
\section{Introduction}
%%%%%%%%%%%%%%%%%%%%%%%%%%%%%%%%%%%%%%%%%%%

Jet vetoes are ubiquitously applied in experimental analyses at the Large Hadron Collider (LHC) to select signal events and suppress the background contribution. 
On the other hand, the jet veto makes the theoretical description of scattering cross sections more complicated, since it constrains the phase space of the QCD radiation to the soft and collinear regions. This induces Sudakov double logarithms in the jet-veto scale at each order in perturbation theory, which need to be resummed to all orders to obtain reliable predictions.

The resummation of these logarithms beyond the common parton-shower accuracy was pioneered in~\cite{Banfi:2012yh} using the {\tt CAESAR} framework. This work triggered a lot of interest in the theory community, and for the prime examples of Drell-Yan and Higgs-boson production, the jet-veto logarithms were subsequently resummed to next-to-next-to-leading logarithmic (NNLL) accuracy, matched to the fixed-order next-to-next-to-leading order (NNLO) prediction~\mbox{\cite{Becher:2012qa,Tackmann:2012bt,Banfi:2012jm,Becher:2013xia,Stewart:2013faa}}. Further refinements that include a resummation of the jet-radius dependence were considered in~\cite{Banfi:2015pju}, while a framework for jointly resumming logarithmic corrections of the Higgs transverse momentum and the jet-veto scale was put forward in~\cite{Monni:2019yyr}. 

The developed formalism is not restricted to Drell-Yan and Higgs-boson production, and a generic framework for resumming jet-veto logarithms for arbitrary electro\-weak final states was developed in~\cite{Becher:2014aya}, while specific analyses of associated Higgs production~\mbox{\cite{Shao:2013uba,Li:2014ria}}, gauge-boson pair production~\cite{Jaiswal:2014yba,Wang:2015mvz,Dawson:2016ysj} and beyond-the-standard-model signatures~\cite{Tackmann:2016jyb,Ebert:2016idf,Fuks:2017vtl,Arpino:2019fmo} were performed at a similar NNLL+NNLO accuracy. Whereas all of these studies used the transverse momentum to veto jets, other proposals that impose a looser constraint in the forward directions~\cite{Gangal:2014qda,Gangal:2020qik} or a rapidity cut~\cite{Hornig:2017pud,Michel:2018hui} were also investigated. Finally, factorisation-breaking corrections from Glauber exchanges were addressed in~\cite{Zeng:2015iba}, where it was argued that these become relevant at N$^4$LL accuracy and beyond.

The theoretical foundation for studying cross sections in the presence of a jet veto can hence be considered as well-established by now. In the language of Soft-Collinear Effective Theory (SCET)~\cite{Bauer:2000yr,Bauer:2001yt,Beneke:2002ph}, the cross section for the production of an electroweak final state of invariant mass $Q$ and rapidity $Y$ with a transverse-momentum veto $\ptv\ll Q$ can be written in the following factorised form~\cite{Becher:2012qa},
\begin{align}
\frac{d^2\sigma(\ptv)}{dQ^2dY} = \sum_{i,j} \, H_{ij}(Q,\mu) \, \mathcal{B}_{i/h_1}(x_1,\ptv,\mu) \, \mathcal{B}_{j/h_2}(x_2,\ptv,\mu) \,\mathcal{S}_{ij}(\ptv,\mu)\,,
\end{align}
where the sum runs over all partonic channels and $x_{1,2}= (Q/\sqrt{s})\,e^{\pm Y}$. Here $H_{ij}(Q,\mu)$ is the hard function that contains the virtual corrections to the Born process, whereas the soft and collinear emissions that pass the jet-veto constraint are described by the soft function $\mathcal{S}_{ij}(\ptv,\mu)$ and the beam functions $\mathcal{B}_{i/h}(x,\ptv,\mu)$, respectively. As long as the jet-veto scale lies in the perturbative regime, $\ptv\gg\Lambda_{\rm QCD}$, the latter can be matched onto the usual parton distribution functions $f_{i/h}(x,\mu)$ via
\begin{align}
\label{eq:matching}
\mathcal{B}_{i/h}(x,\ptv,\mu) &=\sum_k 
    \int_x^1 \frac{dz}{z} \,
    {\cal I}_{i\leftarrow k}\Big(\frac{x}{z},\ptv,\mu\Big) 
    ~ f_{k/h}(z,\mu)\,,
\end{align}
which holds at leading power in $\Lambda_{\rm QCD}/\ptv$. The calculation of the matching kernels ${\cal I}_{i\leftarrow k}(x,\ptv,\mu)$ for the quark beam function with $i=q$ to NNLO accuracy is the goal of the present paper.

The matching kernels for transverse-momentum resummation are, in fact, already known to N$^3$LO accuracy~\cite{Becher:2010tm,Becher:2012yn,Catani:2013tia,Gehrmann:2014yya,Lubbert:2016rku,Echevarria:2016scs,Luo:2019hmp,Luo:2019bmw,Luo:2019szz,Ebert:2020yqt,Luo:2020epw}, as are the ones for N-jettiness~\cite{Stewart:2010qs,Berger:2010xi,Gaunt:2014xga,Gaunt:2014cfa,Melnikov:2018jxb,Melnikov:2019pdm,Behring:2019quf,Baranowski:2020xlp,Ebert:2020unb}. Whereas a double-differential beam function was computed to NNLO in~\cite{Jain:2011iu,Gaunt:2014xxa,Gaunt:2020xlc}, the closely-related beam functions for certain rapidity-dependent jet vetoes are also known at this accuracy~\cite{Gangal:2016kuo}. Somewhat surprisingly, the matching kernels for the standard $\ptv$ are currently only known to NLO~\cite{Becher:2012qa,Shao:2013uba}. While the convolution of two gluon beam functions was numerically extracted from a fixed-order code at NNLO in~\cite{Becher:2013xia,Stewart:2013faa}, no such determination was attempted so far for the quark beam functions. Our calculation therefore provides the ingredients to extend jet-veto resummations for quark-initiated processes to NNLL$'$ accuracy.\footnote{We use the primed-order counting in this article, in which the matching corrections are included at one order higher than in the unprimed one, see for instance Table~6 in~\cite{Bell:2018gce}.}

As the considered jet veto is based on transverse momenta, the observable is described by a version of the effective theory that is referred to as SCET-2. It is well known that perturbative computations in SCET-2 are not well-defined in dimensional regularisation, as they require an additional prescription to regularise rapidity divergences. The matching kernels in \eqref{eq:matching} are therefore regularisation-scheme dependent, and we apply the analytic phase-space regulator proposed in~\cite{Becher:2011dz} in our calculation. As will be described in the following section, the scheme dependence drops out once the collinear, anti-collinear and soft contributions are combined. In order to avoid distribution-valued expressions, we furthermore determine the matching kernels directly in Mellin space.

Our calculation is based on a novel framework that aims at automating the computation of NNLO jet and beam functions for a broad class of observables \cite{Bell:2021dpb,BBDW2}. We indeed checked our numerical predictions against the analytic results for transverse-momentum resummation from~\cite{Gehrmann:2014yya}, before implementing the phase-space constraints that are imposed by the jet veto. The automated setup follows the spirit of {\tt SoftSERVE}~\cite{Bell:2018vaa,Bell:2018oqa,Bell:2020yzz}, which has been successfully applied to compute various NNLO soft functions that involve two light-like Wilson lines. An extension of the soft-function framework to an arbitrary number of jet directions is currently in development~\cite{Bell:2018mkk}.

The outline of this paper is as follows. In Section~\ref{sec:framework} we lay out the theoretical background of our framework, and in Section~\ref{sec:computation} we describe some technical aspects of the NNLO calculation. Our results for the jet-veto matching kernels are presented in Section~\ref{sec:results}, and we conclude in Section~\ref{sec:conclusions}. In the appendix we collect the anomalous dimensions and splitting functions that are needed for the renormalisation-group analysis.

%%%%%%%%%%%%%%%%%%%%%%%%%%%%%%%%%%%%%%%%%%%
\section{Theoretical framework}
%%%%%%%%%%%%%%%%%%%%%%%%%%%%%%%%%%%%%%%%%%%
\label{sec:framework}

We start from the generic definition of a quark beam function,
\begin{align}
\label{eq:definition}
    \frac12  \left[\frac{\ns}{2}\right]_{\beta \alpha}
    {\cal B}_{q/h}(x,\tau,\mu) =& \sum_{X} \,
    \delta\Big( (1-x) P^- - \sum_i k_i^- \Big)\,
		{\cal M}(\tau;\{k_i\}) 
		\nonumber\\
    &\bra{h(P)}\bar{\chi}_{\alpha} \ket{X} 
    \bra{X}\chi_{\beta}  \ket{h(P)} ,
\end{align}
where $\chi = W^{\dagger}_{\bar n}\frac{\ns \nsb}{4} \psi$ is the collinear field operator and $W_{\bar n}$ denotes a collinear Wilson line. We furthermore introduced two light-cone vectors $n^\mu$ and $\bar n^\mu$ satisfying $n^2=\bar n^2=0$ and $n\cdot \bar n =2$, and throughout this article we adopt the notation $k_i^- = \bar n \cdot k_i$, $k_i^+ = n \cdot k_i$ and a transverse component $k_i^{\perp,\mu}$ that fulfils $n \cdot k_i^{\perp}=\bar n \cdot k_i^{\perp}=0$. The sum over $X$ indicates the phase space of the final-state partons with momenta $\{k_i\}$. At tree level this is just the vacuum state, and at NNLO it consists of up to two massless partons. The state $\ket{h(P)}$ refers to a hadronic state of momentum $P^\mu=P^- n^\mu/2$, but in order to extract the matching kernels from the relation \eqref{eq:matching}, it will be convenient to consider partonic states instead. In fact, if the matching is performed on-shell in dimensional regularisation, the parton distribution functions evaluate to $f_{i/j}(x,\mu)=\delta_{ij}\delta(1-x)$ to all orders in perturbation theory, and the calculation of the partonic beam functions directly yields the desired matching kernels.

The above definition of the quark beam function is generic, and the function ${\cal M}(\tau;\{k_i\})$ specifies what is actually measured on the collinear radiation. Following~\cite{Bell:2018oqa,Bell:2020yzz} we write the measurement function in the form
\begin{align}
\mathcal{M}(\tau;\lbrace k_{i} \rbrace) = 
\exp\big(-\tau\, \omega(\lbrace k_{i} \rbrace)\,\big)\,,
\label{eq:measure}
\end{align}
where $\tau$ is a Laplace variable of dimension 1/mass, and the function $\omega(\lbrace k_{i} \rbrace)$ specifies the observable. In the jet-veto case, the phase-space constraints are more naturally formulated in terms of a theta function, $\widehat{\mathcal{M}}(\ptv;\lbrace k_{i} \rbrace) = 
\theta\big(\ptv- \omega(\lbrace k_{i} \rbrace)\,\big)$, and it has been shown in~\cite{Bell:2020yzz} how to convert this into the form \eqref{eq:measure} via a Laplace transform, and how to extract from this calculation the result in the original $\ptv$ space. For one emission with \mbox{momentum $k^\mu$,} the constraint is imposed on the transverse momentum of the emission,
\begin{align}
\label{eq:measure:ptv:nlo}
\omega_1(k) =
|\vec{k}^\perp|\,,
\end{align}
but for two emissions or more the measurement function depends on the distance of the emissions in some predefined measure. If the emissions are close to each other in these units, they are recombined into a pseudo-particle, whereas they are considered as part of two independent jets if they are far away. In the current work we consider the general class of $k_T$-type jet algorithms, and for two emissions one finds that the measurement function depends on the jet radius $R$, but not on the specific clustering prescription according to the $k_T$, Cambridge/Aachen or anti-$k_T$ jet algorithm~\cite{Becher:2012qa}. Denoting the momenta of the two emissions by $k^\mu$ and $l^\mu$, the measurement function can be written in the form
\begin{align}
\label{eq:measure:ptv:nnlo}
\omega_2(k,l) &= 
\theta(\Delta-R)\; \max\big(|\vec{k}^\perp|,|\vec{l}^\perp|\big) + \theta(R-\Delta )\; 
|\vec{k}^\perp+\vec{l}^\perp|\,,
\end{align}
where the distance measure of the jet algorithm translates into
\begin{align}
\Delta= \sqrt{\frac14 \ln^2 \frac{k^-l^+}{k^+l^-}+\theta_{kl}^2}\,,
\end{align}
with $\theta_{kl}$ being the angle between the momenta $\vec{k}^{\perp}$ and $\vec{l}^{\perp}$ in the transverse plane.

Our automated framework for the computation of NNLO beam functions developed in~\cite{Bell:2021dpb,BBDW2} furthermore requires that all distributions are resolved by appropriate integral transformations. For the delta-function constraint in \eqref{eq:definition}, this is achieved by a Mellin transform
\begin{align}
   {\cal \widehat B}_{i/h}(N,\ptv,\mu) = \int_0^1 dx \;
	x^{N-1} \;{\cal B}_{i/h}(x,\ptv,\mu)
	\,,
\end{align}
which brings the matching relation \eqref{eq:matching} into a product form,
\begin{align}
\label{eq:matching:Mellin}
 \mathcal{\widehat B}_{i/h}(N,\ptv,\mu) &=\sum_k 
     \,
    {\cal \widehat I}_{i\leftarrow k}(N,\ptv,\mu) 
    ~ \widehat f_{k/h}(N,\mu)\,.
\end{align}
As the considered jet veto is imposed on the transverse momenta of the reconstructed jets, the relevant soft modes in the effective theory have the same virtuality as the collinear ones. This version of the effective theory is known as SCET-2, and we use the analytic phase-space regulator of~\cite{Becher:2011dz} to regularise  rapidity divergences that are not captured by the dimensional regulator $\eps=(4-d)/2$. Specifically, the rapidity regulator $\alpha$ is introduced on the level of the phase-space measure 
\begin{equation}
\label{eq:regulator}
\int d^dk_i \; \left(\frac{\nu}{k_i^- + k_i^+}\right)^\alpha \;  \delta(k_i^2) \, \theta(k_i^0) 
\end{equation}
for each emitted parton with momentum $k_i^\mu$. While this corresponds to the prescription that is implemented for the soft integrals in {\tt SoftSERVE}, the regulator simplifies in the collinear region with $k_i^- \gg k_i^+$ and in the anti-collinear region with $k_i^+ \gg k_i^-$. As the rapidity regulator respects the $n$-$\bar{n}$ symmetry of the process, it is actually not necessary to compute the anti-quark beam function in the anti-collinear region explicitly.

The rapidity divergences induce a dependence on the rapidity scale $\nu$ that is implicit in \eqref{eq:matching:Mellin}. In order to obtain a result that is independent of the specific regularisation scheme, we follow the collinear-anomaly approach~\cite{Becher:2010tm,Becher:2011pf}, which states that the product of the soft, collinear and anti-collinear functions is finite in the limit $\alpha\to 0$. Specifically, the product of the three functions, which contains an implicit dependence on the hard scale $Q$, can be refactorised in the form
\begin{align}
  &\left[
    \widehat{\cal I}_{q\leftarrow i}(N_1,\ptv,\mu,\nu) \;
    \widehat{\cal I}_{\bar{q}\leftarrow j}(N_2,\ptv,\mu,\nu) \; 
		    {\cal S}_{q\bar{q}}(\ptv,\mu,\nu) 
  \right]_{Q}
	\nonumber\\
  &\qquad=
  \left( \frac{Q}{\ptv}\right)^{-2F_{q\bar{q}}(\ptv,\mu)} \;
  \widehat{I}_{q\leftarrow i}(N_1,\ptv,\mu) \;
  \widehat{I}_{\bar{q}\leftarrow j}(N_2,\ptv,\mu)\,,
	\label{eq:refact}
\end{align}
where the quantities on the right-hand side are manifestly independent of the rapidity \mbox{scale $\nu$}. In this relation, the dependence on the hard scale $Q$ is controlled by the collinear-anomaly exponent $F_{q\bar{q}}(\ptv,\mu)$, which obeys the renormalization-group equation (RGE) 
\begin{align}
\label{eq:anomaly:RGE}
  \frac{d}{d\ln \mu} \; F_{q\bar{q}}(\ptv,\mu) 
  = 2\,\Gamma_{\rm cusp}^F(\als)\,,
\end{align} 
which is controlled by the cusp anomalous dimension in the fundamental representation $\Gamma_{\rm cusp}^F(\als)$. The refactorised matching kernels $\widehat{I}_{i\leftarrow j}(N,\ptv,\mu)$, on the other hand, are independent of the scheme that is used to regularise the rapidity divergences, and their scale dependence is controlled by the RGE
\begin{align}\label{eq:rgeI}
  \frac{d}{d\ln \mu} \;\widehat{I}_{i\leftarrow j}(N,\ptv,\mu)
  =&
  2\left[
    \Gamma_{\rm cusp}^{R_i}(\als) \, L -\gamma^{i}(\als)
  \right] \widehat{I}_{i\leftarrow j}(N,\ptv,\mu)
	\nonumber\\
  &-2 \sum_{k} \,\widehat{I}_{i\leftarrow k}(N,\ptv,\mu) \,\widehat{P}_{k\leftarrow j}(N,\als) \,,
\end{align}
where $L=\ln (\mu/\ptv)$, $\Gamma_{\rm cusp}^{R_i}(\als)$ is the cusp anomalous dimension in the representation of the parton $i$, $\gamma^{i}(\als)$ is the collinear anomalous dimension for quarks ($i=q$) or gluons ($i=g$), and $\widehat{P}_{k\leftarrow j}(N,\als)$ are the DGLAP splitting functions in Mellin space. Explicit solutions to these RGE to the considered two-loop order are given in the following section.

%%%%%%%%%%%%%%%%%%%%%%%%%%%%%%%%%%%%%%%%%%%
\section{Computational aspects}
%%%%%%%%%%%%%%%%%%%%%%%%%%%%%%%%%%%%%%%%%%%
\label{sec:computation}

In our automated setup for computing NNLO beam functions we apply universal phase-space parametrisations to factorise the implicit divergences of the collinear matrix elements. For one emission with momentum $k^\mu$, we use the magnitude of its transverse momentum with respect to the beam axis, which is set by $P^\mu$, and we allow for a non-trivial azimuthal dependence of the observable, i.e.~we choose the variables
\begin{align}
\label{eq:parametrisation:one}
  k_T = |\vec{k}^\perp| \,, 
	\qquad\qquad t_k = \frac{1-\cos \theta_k}{2} \,,
\end{align} 
where $\theta_k$ is the angle between $\vec{k}^\perp$ and a reference vector $\vec{v}^\perp$ that may be imposed by the observable. The remaining momentum components are then fixed by the delta function in \eqref{eq:definition} and the on-shell condition, yielding $k^+=k_T^2/k^-$ with $k^- = (1-x) P^-$.

Following~\cite{Bell:2021dpb,BBDW2} (see also~\cite{Bell:2018vaa,Bell:2018oqa,Bell:2020yzz}), we then parametrise the generic measurement function for a single emission in the form 
\begin{align}
\label{eq:jet:measure:one-emission}
  {\cal M}_1(\tau;k) 
  &= 
  \exp 
  \left[ 
        -\tau k_T 
        \left( 
          \frac{k_T}{(1-x) P^-}
        \right)^n
        f(t_k)
  \right].
\end{align}
One easily verifies that the specific jet-veto measurement from \eqref{eq:measure:ptv:nlo} corresponds to the case $n=0$ and $f(t_k)=1$ in this notation. The above parametrisation is, however, more general and in particular it allows us to gauge our numerical results against the analytic expressions for transverse-momentum resummation from~\cite{Gehrmann:2014yya}, which corresponds to $n=0$ and \mbox{$f(t_k)=2i(1-2t_k)$} in our conventions. The very fact that both of these measurements are described by $n=0$ signals that they are SCET-2 observables~\cite{Bell:2018oqa}. In the latter case, the factor $i$ is moreover a relic of a Fourier transform, and the non-trivial azimuthal dependence arises due to the presence of the electroweak particle in the final state, which singles out a direction in the transverse plane.

For two emissions, we proceed similarly and parametrise the momenta $k^\mu$ and $l^\mu$ of the massless final-state partons in the form
\begin{align}
\label{eq:parametrisation:two}
  a = \frac{k^- l_T}{l^- k_T}, \qquad
  b = \frac{k_T}{l_T}, \qquad
  z = \frac{k^- + l^-}{P^-}, \qquad
  q_T = \sqrt{(k^- + l^-)(k^+ + l^+)}\,,
\end{align}
where again $k_T = |\vec{k}^\perp|$ and $l_T = |\vec{l}^\perp|$,
and in addition we now have three non-trivial angles in the transverse plane, $\theta_k$ and $\theta_l$ which refer to the reference vector $\vec{v}^\perp$ as before, and $\theta_{kl}$ which is the angle between $\vec{k}^\perp$ and $\vec{l}^\perp$. We then rewrite these angles in terms of the variables $t_k$, $t_l$ and $t_{kl}$ similar to \eqref{eq:parametrisation:one}, which are defined on the unit interval.

In terms of these variables, we make the following ansatz for the two-emission measure\-ment function,
\begin{align}
  {\cal M}_2(\tau; k,l) 
  &= 
  \exp 
  \left[ 
        -\tau q_T 
        \left( 
          \frac{q_T}{(1-x) P^-}
        \right)^n
        {\cal F}(a,b,z,t_k, t_l, t_{kl})
  \right],
\end{align}
which captures the jet-veto case from \eqref{eq:measure:ptv:nnlo}
 for 
\begin{align}
 & {\cal F}(a,b,z,t_k, t_l, t_{kl}) =
\rho\,\bigg\{
	\theta(\Delta-R)\; \max(1,b) + \theta(R-\Delta )\;
	\sqrt{(1 - b)^2 + 4 b (1 - t_{kl})} \bigg\}\,,
\end{align}
with $\rho=\sqrt{a/(1+a b)/(a+b)}$ and $\Delta= \sqrt{\ln^2 a+\arccos^2(1-2t_{kl})}$, while the measure for transverse-momentum resummation is given by ${\cal F}(a,b,z,t_k, t_l, t_{kl}) = 2i\rho\big(b(1-2t_k)+1-2t_l\big)$. The very fact that the latter involves a Fourier rather than a Laplace transform requires a workaround in our numerical approach that was described in detail in the appendix of \cite{Bell:2018oqa}.

Having fixed the phase-space parametrisations and the form of the measurement function, the partonic beam functions can readily be evaluated. The required collinear matrix elements are known to be related to spin-averaged $d$-dimensional splitting functions after crossing~\cite{Ritzmann:2014mka}. Specifically, we use the expressions from~\cite{Kosower:1999rx,Bern:1999ry,Sborlini:2013jba} for the real-virtual contribution, and from~\cite{Campbell:1997hg,Catani:1998nv} for the double real-emission part at NNLO.
While the phase-space divergences of the former can easily be factorised in terms of the variables given in \eqref{eq:parametrisation:one}, this is only true in certain cases for the latter using \eqref{eq:parametrisation:two}. Several further steps that involve sector-decomposition techniques, non-linear transformations and selector functions are, in fact, needed to factorise all divergences of the double real-emission contribution. Once this factorisation is accomplished, it becomes a straightforward task to perform a Laurent expansion in the rapidity regulator $\alpha$ and the dimensional regulator $\eps$, and we finally integrate the respective coefficients in this double expansion numerically. To perform these steps, we have implemented our formalism in the public program {\tt pySecDec}~\cite{Borowka:2017idc}, and we use the {\tt Vegas} routine of the {\tt Cuba} library~\cite{Hahn:2004fe} for the numerical integrations.

By following this procedure, which will be described in more detail in~\cite{BBDW}, one obtains the regulator-dependent  matching kernels on the left-hand side of the refactorisation condition \eqref{eq:refact}. In order to extract the regulator-independent matching kernels on the right-hand side of this equation, one needs to combine the collinear and anti-collinear matching kernels with the corresponding soft function in the same regularisation scheme. To this end, we use   {\tt SoftSERVE}, which contains both the considered jet veto and transverse-momentum resummation among the template observables that are included in the distribution. As a first check of our calculation, we then observe that the $1/\alpha$ poles and the dependence on the rapidity scale $\nu$ drop out in the product of the three functions, and we extract the bare anomaly exponent and the bare refactorised matching kernels using \eqref{eq:refact}. The former is known to renormalise additively, $F_{q\bar{q}}^{\rm bare}=F_{q\bar{q}}+Z_{q\bar{q}}^F$, and the solution to its RGE \eqref{eq:anomaly:RGE} up to the considered two-loop order is given by
\begin{align}
  F_{q\bar{q}}(\ptv,\mu)  &= 
\left( \frac{\alpha_s}{4 \pi} \right) 
\Big\{ 2\Gamma_0^F \,L 
+ d_1 \Big\}
+\left( \frac{\alpha_s}{4 \pi} \right)^2 
\Big\{ 2 \beta_0\Gamma_0^F\, L^2 
 + 2 \left( \Gamma_1^F + \beta_0 d_1 \right) L + d_2 \Big\},
\end{align}
where $\Gamma_i^F$ and $\beta_i$ are the expansion coefficients of the cusp anomalous dimension and the $\beta$-function, respectively, in the conventions that are specified in Appendix~\ref{app:AD}. The non-logarithmic terms $d_i$ of the anomaly exponent are, in fact, also known to the considered two-loop order for both transverse-momentum~\cite{Becher:2010tm} and jet-veto~\cite{Banfi:2012yh,Becher:2013xia,Stewart:2013faa} resummation, and we can hence use these expressions as a further cross-check of our calculation. For completeness, we also provide the two-loop solution of the counterterm $Z_{q\bar{q}}^F$, which obeys a similar RGE as the renormalised anomaly exponent. It reads
\begin{align} 
Z_{q\bar{q}}^F(\ptv,\mu)&= \left( \frac{\alpha_s}{4 \pi} \right) 
\left\{ \frac{\Gamma_0^F}{\eps} \right\}
+\left( \frac{\alpha_s}{4 \pi} \right)^2 
\bigg\{ - \frac{\beta_0\Gamma_0^F}{2\eps^2}
+ \frac{\Gamma_1^F}{2\eps}
\bigg\}\,.
\end{align}
We next consider the refactorised matching kernels that are defined via the relation \eqref{eq:refact}. In Mellin space, these kernels renormalise multiplicatively, but since they contain two different sources of divergences, we find it convenient to introduce two types of counterterms in this case. Explicitly, we write $\widehat{I}_{i\leftarrow j} = Z_i^B \sum_k\widehat{I}_{i\leftarrow k}^{\rm bare} \,\widehat{Z}_{k\leftarrow j}^f$,
where $Z_i^B$ captures the UV divergences of the beam function defined in \eqref{eq:definition} (for $i=q$), whereas $\widehat{Z}_{k\leftarrow j}^f$ subtracts the IR divergences that match the UV divergences of the parton distribution functions. The former counterterm obeys the RGE
\begin{align}
  \frac{d}{d\ln \mu} \;
	Z_i^B(\ptv,\mu)
  =&
  2\left[
    \Gamma_{\rm cusp}^{R_i}(\als) \, L -\gamma^{i}(\als)
  \right] Z_i^B(\ptv,\mu) \,,
\end{align}
which to two-loop order is solved by
\begin{align}
&Z_i^B(\ptv,\mu)  =  
1 + \left( \frac{\alpha_s}{4 \pi} \right) 
\left\{ - \frac{\Gamma_0^i}{2\eps^2} - 
\frac{\Gamma_0^i L - \gamma_0^i}{\eps} 
\right\}
\nonumber\\[0.2em]  
 &\quad
+\left( \frac{\alpha_s}{4 \pi} \right)^2 
\bigg\{ \frac{(\Gamma_0^i)^2}{8\eps^4}
+ \left( \frac{\Gamma_0^i}{2}L
-\frac{\gamma_0^i}{2}
 + \frac{3\beta_0}{8}\right) \frac{\Gamma_0^i}{\eps^3}
+\bigg( \frac{(\Gamma_0^i)^2}{2}L^2 
- \Gamma_0^i \Big(
\gamma_0^i - \frac{\beta_0}{2} \Big)L
\nonumber\\[0.2em]  
 &\qquad\qquad\qquad
-\frac{\Gamma_1^i}{8} 
+ \frac{(\gamma_0^i)^2}{2}
- \frac{\beta_0\gamma_0^i}{2}
\bigg) \, \frac{1}{\eps^2}
 - \frac{\Gamma_1^i L - \gamma_1^i}{2\eps}
\bigg\}
\,,
\end{align}
and the corresponding coefficients of the anomalous dimensions can again be found in Appendix~\ref{app:AD}. 

While the beam-function counterterm $Z_i^B$ subtracts double Sudakov-type divergences that are controlled by the cusp anomalous dimension, the parton distribution functions are known to have a single-logarithmic evolution. The non-trivial aspect, on the other hand, is in this case the operator mixing. The RGE of the respective counterterm reads
\begin{align}
  \frac{d}{d\ln \mu} \;
	\widehat{Z}_{k\leftarrow j}^f(N,\ptv,\mu)
  =&
  -2 \sum_{l}  \,\widehat{Z}_{k\leftarrow l}^f(N,\ptv,\mu)
	\,\widehat{P}_{l\leftarrow j}(N,\als)\,,
\end{align}
where $\widehat{P}_{l\leftarrow j}(N,\als)$ are the Mellin-space DGLAP splitting functions that we introduced earlier in \eqref{eq:rgeI}. The general solution to this RGE is given by
\begin{align}
&\widehat{Z}_{k\leftarrow j}^f(N,\ptv,\mu)  =  
\delta_{kj} + \left( \frac{\alpha_s}{4 \pi} \right) 
\left\{ \widehat{P}_{k\leftarrow j}^{(0)}(N)\,\frac{1}{\eps}
\right\}
\nonumber\\[0.2em]  
&\quad
+\left( \frac{\alpha_s}{4 \pi} \right)^2 
\bigg\{ 
- \widehat{P}_{k\leftarrow j}^{(0)}(N)\,
\frac{\beta_0}{2\eps^2}
+\sum_l \, \widehat{P}_{k\leftarrow l}^{(0)}(N) \
 \widehat{P}_{l\leftarrow j}^{(0)}(N)  \, \frac{1}{2\eps^2}
+ \widehat{P}_{k\leftarrow j}^{(1)}(N)\,\frac{1}{2\eps}\bigg\}
\,,
\end{align}
where the $\widehat{P}_{i\leftarrow j}^{(m)}(N)$ are the $m$-th order coefficients of the DGLAP splitting functions that are also defined in Appendix~\ref{app:AD}.

After combining the bare refactorised matching kernels with these counterterms as described above, we find that all $1/\eps$ divergences drop out to the numerical accuracy of our calculation. We then extract the renormalised matching kernels, which up to two-loop order can be written in the form
\begin{align}
\label{eq:Iren}
\widehat{I}_{i\leftarrow j}&(N,\ptv,\mu)  
\\[0.2em]  
= &\;
\delta_{ij} + \left( \frac{\alpha_s}{4 \pi} \right) 
\bigg\{ \Big( \Gamma_0^i \,L^2 
- 2\gamma_0^i \,L \Big) \delta_{ij}  
- 2L\,\widehat{P}_{i\leftarrow j}^{(0)}(N)
+ \widehat{I}_{i\leftarrow j}^{(1)}(N) \bigg\}
\nonumber\\[0.2em]  
  &\;
+\left( \frac{\alpha_s}{4 \pi} \right)^2 
\bigg\{ \bigg( \frac{(\Gamma_0^i)^2}{2} L^4 
-  2\Gamma_0^i\left( \!\gamma_0^i - \frac{\beta_0}{3} \right) 
L^3 + \big( \Gamma_1^i + 2(\gamma_0^i)^2 
 - 2\beta_0 \gamma_0^i  \big) L^2  
-  2\gamma_1^i L\bigg) \delta_{ij}
\nonumber\\[0.2em]  
 &\qquad\qquad\quad
-2\Big( \Gamma_0^i L^3 + \big( \beta_0 - 2 \gamma_0^i\big) L^2\Big) \widehat{P}_{i\leftarrow j}^{(0)}(N)
 + \Big( \Gamma_0^i  L^2 -  2(\gamma_0^i -\beta_0) L \Big)\,
\widehat{I}_{i\leftarrow j}^{(1)}(N)  
\nonumber\\[0.2em]  
 &\qquad\qquad\quad
 +2\sum_k \bigg( L^2\, \widehat{P}_{i\leftarrow k}^{(0)}(N) 
- L \, \widehat{I}_{i\leftarrow k}^{(1)}(N) \bigg) \widehat{P}_{k\leftarrow j}^{(0)}(N)  
- 2L\,\widehat{P}_{i\leftarrow j}^{(1)}(N)
+  \widehat{I}_{i\leftarrow j}^{(2)}(N) \bigg\} .
\nonumber
\end{align}
While the logarithmic terms in this expression are controlled by known anomalous dimensions and splitting functions, the goal of the present article consists in determining the non-logarithmic terms $\widehat{I}_{i\leftarrow j}^{(m)}(N)$. We in fact use the known analytic results for transverse-momentum resummation as a final cross-check of our calculation, and we provide the two-loop coefficients $\widehat{I}_{q\leftarrow j}^{(2)}(N)$ for jet-veto resummation for the first time in this work.

%%%%%%%%%%%%%%%%%%%%%%%%%%%%%%%%%%%%%%
\section{Results}
%%%%%%%%%%%%%%%%%%%%%%%%%%%%%%%%%%%%%%
\label{sec:results}

We now present our results for the renormalised matching kernels $\widehat{I}_{q\leftarrow j}^{(m)}(N)$ up to two-loop order. As QCD is invariant under charge conjugation, the corresponding anti-quark kernels $\widehat{I}_{\bar q\leftarrow j}^{(m)}(N)$ can be directly derived from these expressions. Although we obtained similar results for transverse-momentum resummation with our generic setup, we only show the jet-veto kernels here, since the former are already known for a long time and they serve only as a cross-check of our calculation. Our results for transverse-momentum resummation have, in fact, already been presented in~\cite{Bell:2021dpb,BBDW2}. 

Whereas the matching kernels are trivial at tree level, see \eqref{eq:Iren}, they can easily be evaluated analytically at NLO, since the complicated phase-space constraints that are imposed by the jet algorithm are still ineffective at this order. In Mellin space, the kernels for the two non-vanishing channels at this order are given by
\begin{align}
\widehat{I}_{q\leftarrow q}^{(1)}(N)&=
C_F\bigg(\frac{2}{N} - \frac{2}{N + 1} - \frac{\pi^2}{6}\bigg) \,,
\nonumber\\
\widehat{I}_{q\leftarrow g}^{(1)}(N) &=
T_F\bigg(\frac{4}{N + 1} - \frac{4}{N + 2}\bigg)\,,
\end{align}
whereas the one-loop anomaly exponent vanishes, $d_1=0$.

\begin{figure}[t!]
    \centerline{
    \includegraphics[width=0.45\textwidth]{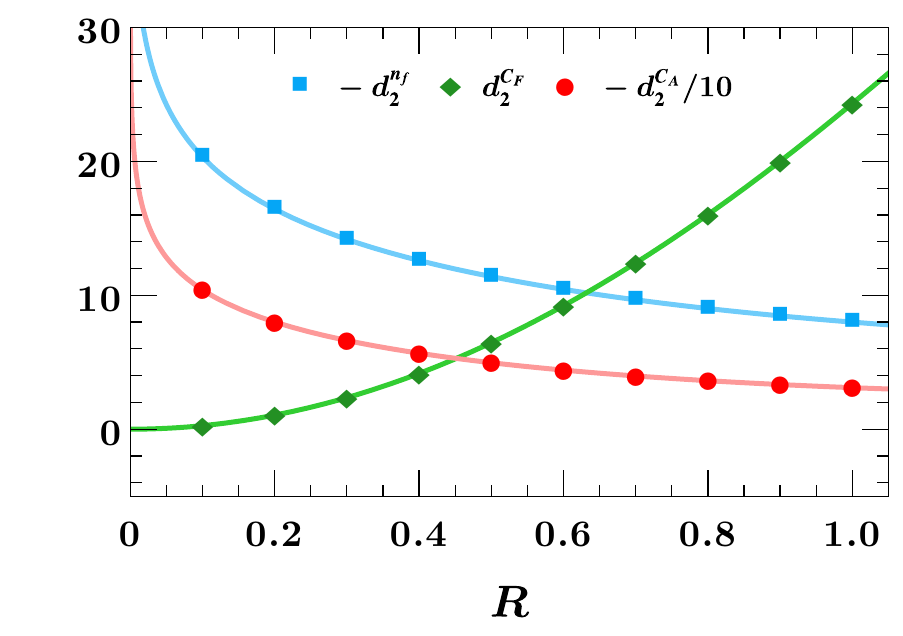}
    \includegraphics[width=0.45\textwidth]{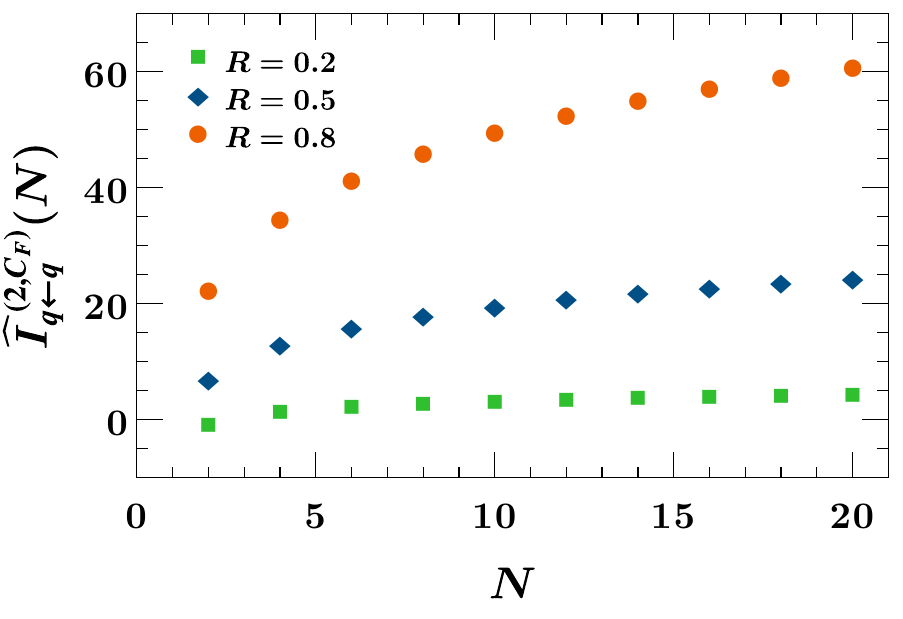}
    }
%   \vspace{3mm}
  \centerline{
    \includegraphics[width=0.45\textwidth]{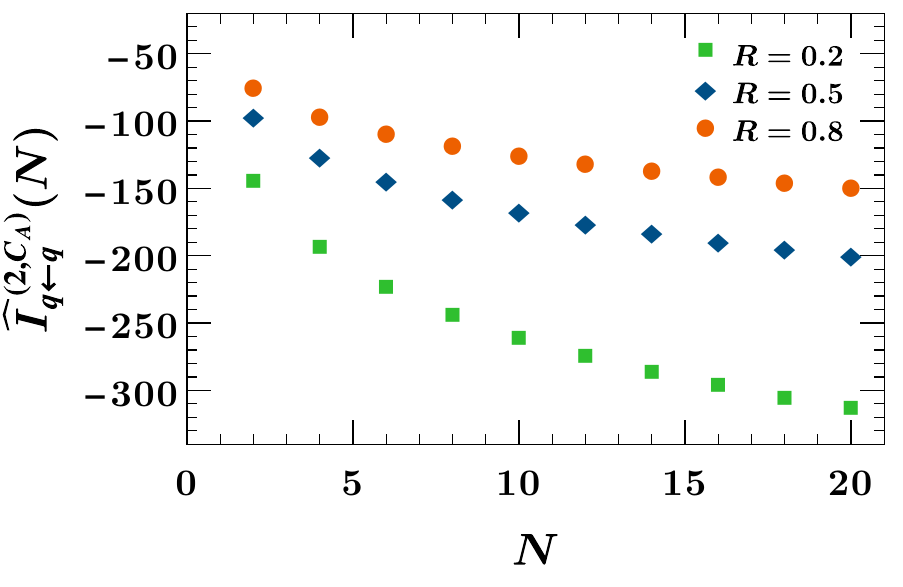}
    \includegraphics[width=0.45\textwidth]{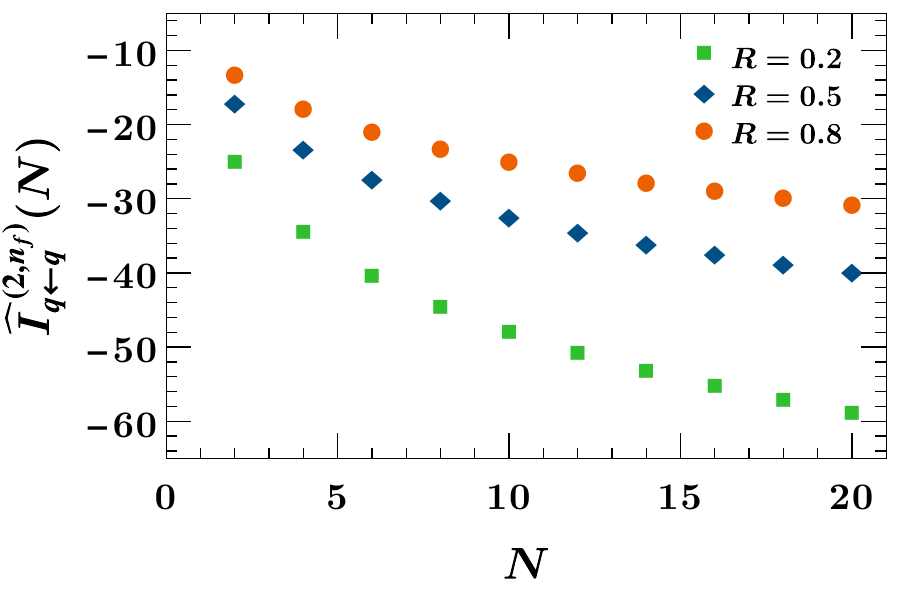}
    }
    \vspace{-2mm}
    \caption{\small{Upper left: Two-loop anomaly exponent $d_2$ as a function of the jet radius $R$. The dots show the results of our numerical evaluation, and the lines represent the expansion from~\cite{Becher:2013xia}.
Other: Non-logarithmic terms of the NNLO matching kernels defined in \eqref{eq:NNLOkernels} for three different values of the jet radius $R$.
}}
\label{fig:anomaly}
\end{figure}

At NNLO we sample the matching kernels for ten values of the Mellin parameter $N\in\{2,4,6,8,10,12,14,16,18,20\}$ and three values of the jet radius $R\in\{0.2,0.5,0.8\}$. In order to gauge the accuracy of our numerical predictions, we first extract the known two-loop quark anomalous dimension,
\begin{align} 
\gamma_1^q &= \gamma_1^{C_F} \,C_F^2
+ \gamma_1^{C_A} \,C_F C_A 
+ \gamma_1^{n_f} \,C_F T_F n_f\,,
\end{align}
and the two-loop anomaly exponent,
\begin{align} 
d_2 &= d_2^{C_F} \,C_F^2
+ d_2^{C_A} \,C_F C_A 
+ d_2^{n_f} \,C_F T_F n_f\,,
\end{align}
from our calculation. Specifically, we use our numbers for the quark-to-quark matching kernel with $N=8$ and $R=0.5$ for this purpose. For the quark anomalous dimension, we obtain
\begin{alignat}{2}
\gamma_1^{C_F}&= -10.6105 (152) \, &&\qquad
[- 10.6102]\,, 
\no\\
\gamma_1^{C_A}&= -4.6424 (142)\, &&\qquad
[-4.6371]\,, 
\no\\
\gamma_1^{n_f}&= 11.3948 (20)\, &&\qquad
[11.3946]\,, 
\end{alignat}
where the known analytic numbers are shown in the square brackets for comparison. The two-loop anomaly exponent, on the other hand, depends on the jet radius $R$ and is displayed in the upper left panel of Figure~\ref{fig:anomaly}. In this plot, we also show the known results from~\cite{Becher:2013xia}, which were obtained semi-analytically in an expansion for $R\ll 1$. Note that our results have numerical uncertainties, which are not visible on the scale of the plot, while the results from~\cite{Becher:2013xia} have an inherent error due to the truncation of the small-$R$ expansion. Overall, the plot shows that the two-loop anomaly is correctly reproduced by our calculation, and that the expansion used in~\cite{Becher:2013xia} extends up to large values of $R\lesssim1$.

\begin{figure}[t!]
    \centerline{
    \includegraphics[width=7.5cm, height=5.0cm]{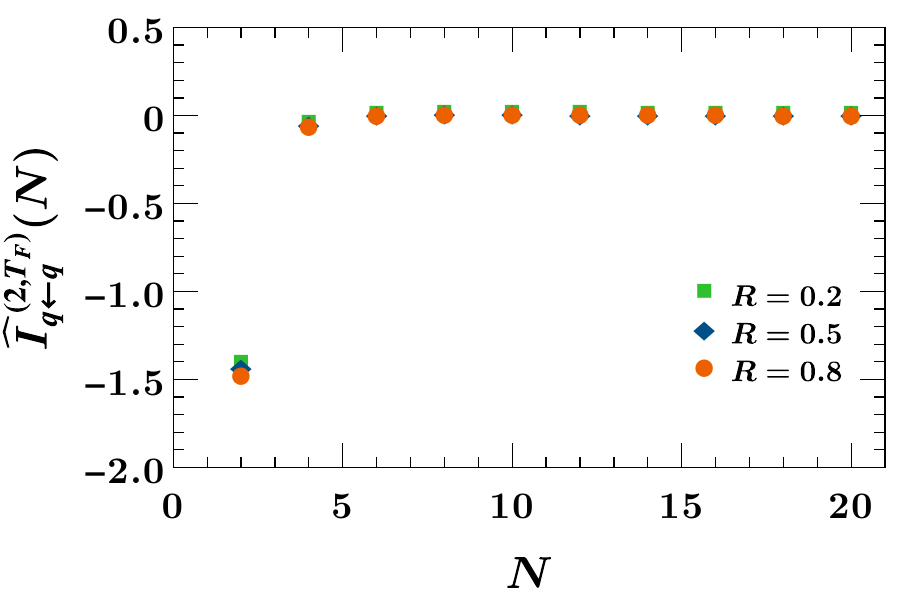}
    \includegraphics[width=7.5cm, height=5.0cm]{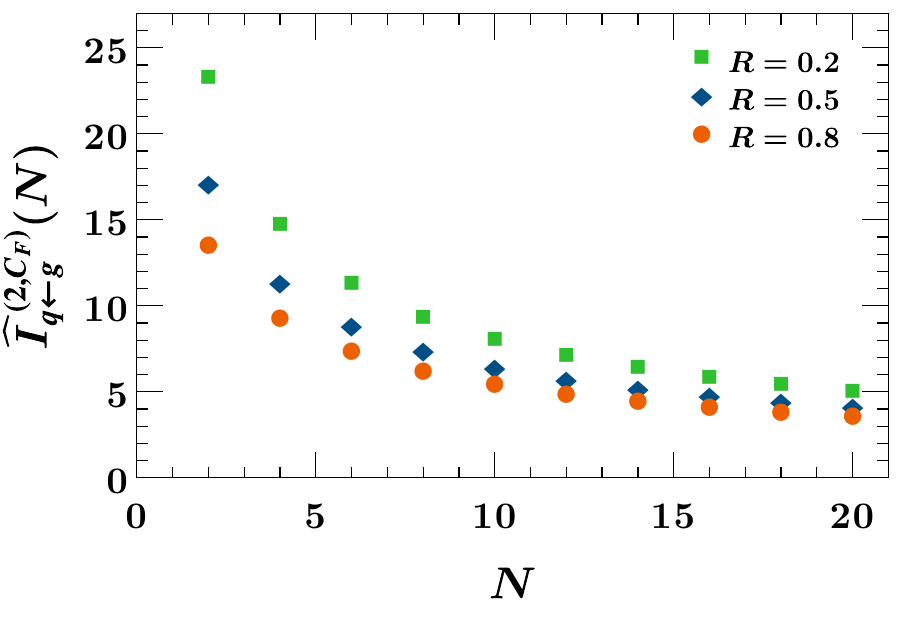}
    }
    \vspace{3mm}
    \centerline{
    \includegraphics[width=7.5cm, height=5.0cm]{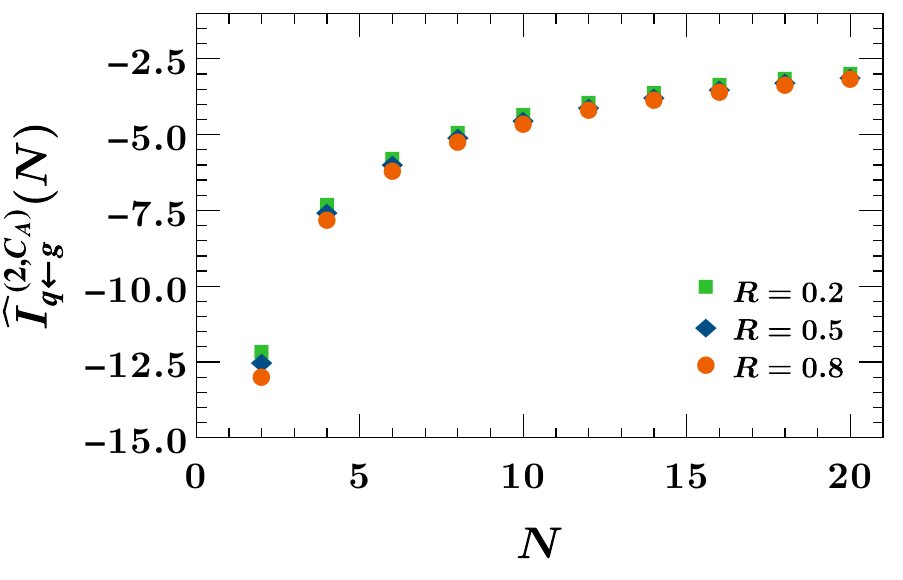}
    \includegraphics[width=7.5cm, height=5.0cm]{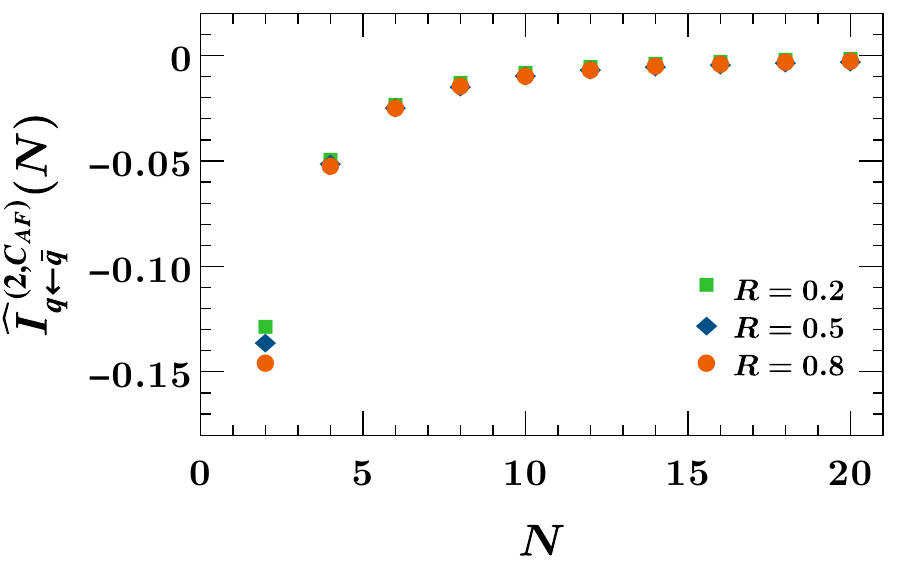}
				}
    \vspace{-2mm}
    \caption{\small{Non-logarithmic terms of the NNLO matching kernels defined in \eqref{eq:NNLOkernels}.
   }}
\label{fig:NNLOkernels}
\end{figure}

For the non-logarithmic terms of the matching kernels we use the decomposition
\begin{align}
\widehat{I}_{q\leftarrow q}^{(2)}(N)&=
C_F^2 \; \widehat{I}_{q\leftarrow q}^{(2,C_F)}(N)
+ C_F C_A \; \widehat{I}_{q\leftarrow q}^{(2,C_A)}(N)   
+ C_F T_F n_f \; \widehat{I}_{q\leftarrow q}^{(2,n_f)}(N)
+ C_F T_F \; \widehat{I}_{q\leftarrow q}^{(2,T_F)}(N)\,,
\nonumber\\
\widehat{I}_{q\leftarrow g}^{(2)}(N) &=
C_F T_F \; \widehat{I}_{q\leftarrow g}^{(2,C_F)}(N) 
+ C_A T_F \; \widehat{I}_{q\leftarrow g}^{(2,C_A)}(N) \,,
\nonumber\\
\widehat{I}_{q\leftarrow \bar q}^{(2)}(N) &=
C_F (C_A - 2 C_F) \; \widehat{I}_{q\leftarrow \bar q}^{(2,C_{AF})}(N)  
+ C_F T_F \; \widehat{I}_{q\leftarrow q}^{(2,T_F)}(N) \,,
\nonumber\\
\widehat{I}_{q\leftarrow q'}^{(2)}(N) &=
\widehat{I}_{q\leftarrow \bar q'}^{(2)}(N) =
C_F T_F \; \widehat{I}_{q\leftarrow q}^{(2,T_F)}(N)\,.
\label{eq:NNLOkernels}
\end{align}
Apart from the quark-to-quark and gluon-to-quark channels that are already present at NLO, there are further channels opening up at NNLO, and we write $q'$ and $\bar q'$ to indicate that the initiating parton has a different flavour than the quark that appears in the definition of the beam function. In total, there are thus seven independent matching kernels at this order that we resolve numerically in Mellin space in our approach. The first three quark-to-quark kernels are shown in Figure~\ref{fig:anomaly}, and the remaining four kernels are displayed in Figure~\ref{fig:NNLOkernels}.

We emphasise again that our results have numerical uncertainties, which can be reconstructed from the electronic file that accompanies the present article, but which are not visible on the scale of the plot. Except in special cases, in which the central value by itself is very small, these uncertainties are well below the percent level, and they are, in fact, smaller for the jet-veto kernels than for the ones of transverse-momentum resummation, for which the Fourier transform is known to pose numerical challenges in our framework~\cite{Bell:2018oqa}. Interestingly, for some kernels the dependence on the jet radius $R$ is quite pronounced, while it is less so for others. 
While we focused here on real Mellin values for illustration purposes, we emphasise that our method can, moreover, equally be applied for complex values of $N$, as required for inverting the Mellin transform.

%%%%%%%%%%%%%%%%%%%%% Remainder coefficient nf

%%%%%%%%%%%%%%%%%%%%
\section{Conclusion}
%%%%%%%%%%%%%%%%%%%%
\label{sec:conclusions} 

We computed the beam-function matching kernels for jet-veto resummation in quark-initiated processes to NNLO. As the considered jet veto  is imposed on the transverse momenta of the reconstructed jets, the beam function is defined in SCET-2, and we used the analytic phase-space regulator of~\cite{Becher:2011dz} to regularise rapidity divergences. We furthermore combined the matching kernels with the corresponding soft function, which we calculated with {\tt SoftSERVE}, to obtain scheme-independent expressions. Our final results for the Mellin-space matching kernels are shown in Figures~\ref{fig:anomaly} and \ref{fig:NNLOkernels}, and they are also provided in electronic form as supplementary material attached to this paper.

While we focused on real values of the Mellin parameter $N$ for illustration, our method is not restricted to this assumption. One may thus use our method to numerically invert the Mellin transform by choosing an appropriate contour in the complex Mellin plane. While this requires a careful treatment of the distributional terms in momentum space, the run-time of our codes is small enough to use very fine grids for this inversion, which guarantees that the associated systematic uncertainties are under control. In a future work, we plan to  present an alternative method for computing beam-function matching kernels directly in momentum space.

Our computation is based on a novel framework that automates the calculation of NNLO beam functions~\cite{Bell:2021dpb}. Apart from the transverse-momentum and jet-veto beam functions considered in this work, we in fact already applied this formalism to other SCET-1 and SCET-2 observables~\cite{BBDW2}. For the jet veto, our calculation provides the ingredients to extend the resummations for quark-initiated processes to NNLL$'$ accuracy. It would furthermore be interesting to derive the respective gluon beam function with the same method, which we leave for future work. In the long term, we plan to provide a public code for the computation of NNLO beam functions in the spirit of the {\tt SoftSERVE} distribution.\\

\noindent
\textbf{Note added:}
Shortly after our work was published on arXiv, Ref.~\cite{Abreu:2022zgo} appeared which also addresses the NNLO jet-veto matching kernels. While the authors of~\cite{Abreu:2022zgo} in addition considered the gluonic channels, their computational approach seems very different from ours, as they used, for instance, a different rapidity regulator, they only computed the difference with respect to a reference observable, and they calculated the matching kernels directly in momentum space. Our approach, on the other hand, provides a straight-forward calculation in Mellin space that does not rely on any reference observable, and it can therefore easily be adapted to other observables.

\acknowledgments
We thank Ding Yu Shao for discussions. This work was supported by the Deutsche Forschungsgemeinschaft (DFG, German Research Foundation) under grant 396021762 - TRR 257 (\emph{``Particle Physics Phenomenology after the Higgs Discovery''}). G.B.~thanks the Department of Fundamental Physics at the University of Salamanca for hospitality.

\begin{appendix}

\section{Anomalous dimensions and splitting functions}
\label{app:AD}

We define the coefficients in the perturbative expansion of the anomalous dimensions as
\begin{align}
\Gamma_{\mathrm{cusp}}^R(\alpha_s) = \sum_{m=0}^{\infty} \left(\frac{\alpha_{s}}{4 \pi} \right)^{m+1} \Gamma_{m}^R\,, \qquad 
\gamma^{i}(\alpha_s)= \sum_{m=0}^{\infty} \left(\frac{\alpha_{s}}{4 \pi} \right)^{m+1} \gamma^{i}_{m}\,,
\end{align} 
and similarly for the QCD $\beta$-function
\begin{align}
\beta(\alpha_s) = -2\alpha_s \sum_{m=0}^{\infty} \left(\frac{\alpha_{s}}{4 \pi} \right)^{m+1} \beta_{m}\,.
\end{align} 
In this work we need the cusp anomalous dimension in the fundamental representation and the collinear quark anomalous dimension to two-loop order~\cite{Becher:2006mr},
\begin{align} 
\Gamma_0^F &= 4 C_F\,,
\\
\Gamma_1^F &= 4 C_F
\left\{ \left(\frac{67}{9}-\frac{\pi^2}{3}\right) C_A
- \frac{20}{9} T_F n_f \right\}\,,
\nonumber\\
\gamma_0^q &= -3C_F\,,
\nonumber\\
\gamma_1^q &=C_F^2 \bigg( 2\pi^2 -\frac32-24\zeta_3\bigg)
+ C_F C_A \bigg(26\zeta_3-\frac{961}{54} - \frac{11\pi^2}{6} \bigg) + C_F T_F n_f \bigg(\frac{130}{27}+\frac{2\pi^2}{3} \bigg)\,,
\nonumber
\end{align}
whereas the $\beta$-function is only required to one-loop order,
\begin{align} 
\qquad
\beta_0 = \frac{11}{3} C_A - \frac43T_F n_f\,.
\end{align} 
We furthermore expand the Mellin-space splitting functions in the form
\begin{align}
\widehat{P}_{j\leftarrow i}(N,\als)=
\sum_{m=0}^{\infty} \left(\frac{\alpha_{s}}{4 \pi} \right)^{m+1} 
\widehat{P}_{j\leftarrow i}^{(m)}(N)\,,
\end{align} 
which at one-loop order are given by
\begin{align}
\widehat{P}_{q\leftarrow q}^{(0)}(N)=&
C_F \bigg(3 - 2 \mathcal{H}_{N + 1} - 2 \mathcal{H}_{N}
+ \frac{2}{N} \bigg)\,,
\nonumber\\
\widehat{P}_{q\leftarrow g}^{(0)}(N)=&
T_F \bigg( \frac{4}{N + 2} - \frac{4}{N + 1} + \frac{2}{N} \bigg)\,,
\nonumber\\
\widehat{P}_{g\leftarrow q}^{(0)}(N)=&
C_F \bigg( \frac{2}{N + 1} - \frac{4}{N} + \frac{4}{N - 1} \bigg)\,,
\nonumber\\
\widehat{P}_{g\leftarrow g}^{(0)}(N)=&
C_A \bigg(-4 \mathcal{H}_{N - 2} - \frac{4}{N + 2} + \frac{4}{N + 1} - \frac{8}{N} \bigg) + \beta_0\,,
\end{align} 
where $\mathcal{H}_{N}$ is the $N$-th harmonic number. At two-loop order and beyond, the Mellin-space splitting functions are often expressed in terms of harmonic sums 
(see e.g. \cite{Moch:2004pa,Vogt:2004mw,Blumlein:2021enk}), but we do not repeat these lengthy expressions here. Instead we provide numerical values for the Mellin parameters that we used in our analysis in Section~\ref{sec:results}. To this end, we start from a similar decomposition as in \eqref{eq:NNLOkernels}
\begin{align}
\widehat{P}_{q\leftarrow q}^{(1)}(N)&=
C_F^2 \; \widehat{P}_{q\leftarrow q}^{(1,C_F)}(N)
+ C_F C_A \; \widehat{P}_{q\leftarrow q}^{(1,C_A)}(N)   
+ C_F T_F n_f \; \widehat{P}_{q\leftarrow q}^{(1,n_f)}(N) 
+ C_F T_F \; \widehat{P}_{q\leftarrow q}^{(1,T_F)}(N)\,,
\nonumber\\
\widehat{P}_{q\leftarrow g}^{(1)}(N) &=
C_F T_F \; \widehat{P}_{q\leftarrow g}^{(1,C_F)}(N) 
+ C_A T_F \; \widehat{P}_{q\leftarrow g}^{(1,C_A)}(N) \,,
\nonumber\\
\widehat{P}_{q\leftarrow \bar q}^{(1)}(N) &=
C_F (C_A - 2 C_F) \; \widehat{P}_{q\leftarrow \bar q}^{(1,C_{AF})}(N) + C_F T_F \; \widehat{P}_{q\leftarrow q}^{(1,T_F)}(N) \,,
\nonumber\\
\widehat{P}_{q\leftarrow q'}^{(1)}(N) &=
\widehat{P}_{q\leftarrow \bar q'}^{(1)}(N) =
C_F T_F \; \widehat{P}_{q\leftarrow q}^{(1,T_F)}(N)\,.
\end{align}
The numerical values we obtained for these coefficients 
by integrating the momentum-space splitting functions 
provided in~\cite{Curci:1980uw,Furmanski:1980cm,Ellis:1996nn} are shown in Table~\ref{tab:splittingfunctions:two-loop}.

\begin{table}[!ht]
    \centering
    \begin{tabular}{|c|c|c|c|c|c|c|c|}
    \hline
        $N$ &
        $\widehat{P}_{q\leftarrow q}^{(1,C_F)}$ &
        $\widehat{P}_{q\leftarrow q}^{(1,C_A)}$ &
				$\widehat{P}_{q\leftarrow q}^{(1,n_f)}$ &
				$\widehat{P}_{q\leftarrow q}^{(1,T_F)}$ &
				$\widehat{P}_{q\leftarrow g}^{(1,C_F)}$ &
				$\widehat{P}_{q\leftarrow g}^{(1,C_A)}$ &
				$\widehat{P}_{q\leftarrow \bar q}^{(1,C_{AF})}$
				\\
    \hline
    \hline
        $2$ &
				  3.9979&
					-13.8508 &
					4.7407		&
					1.4815&
					2.7407&
					1.2963&
					-0.0751
					\\
    \hline
		        $4$ &
				 5.3153 &
          -23.9337&
					9.8304&
					0.0795&
					3.1287&
					-1.8208&
					-0.0026\\
    \hline
        $6$ &
				 6.1672 &
          -29.7455&
					12.9439&
					0.0156&
					3.1971&
					-2.3413&
					-0.0003\\
    \hline
        $8$ &
				6.7588 &
          -33.9480&
					15.2188&
					0.0049&
					3.1287&
					-2.4719&
					0.0000\\
    \hline
        $10$ & 	
				7.1955 &
				-37.2665& 
					17.0189&
					0.0020&
					3.0170&
					-2.4766&
					0.0000\\
    \hline
        $12$ &
				 7.5327 &
          -40.0172&
					18.5109&
					0.0010&
					2.8952&
					-2.4330&
					0.0000\\
    \hline
        $14$ &
				7.8021 &
          -42.3694&
					19.7860&
					0.0005&
					2.7755&
					-2.3702&
					0.0000\\
    \hline
        $16$ &
				8.0228 &
				-44.4256&
					20.8997&
					0.0003&
					2.6624&
					-2.3007&
					0.0000\\
    \hline
        $18$ &
				8.2074 &
			-46.2527&
					21.8885&
					0.0002&
					2.5573&
					-2.2302&
					0.0000\\
    \hline
        $20$ &
				 8.3645 &
          -47.8972&
					22.7777&
					0.0001&
					2.4602&
					-2.1613&
					0.0000\\
    \hline
    \end{tabular}
\caption{Two-loop splitting functions evaluated at sample values of the Mellin parameter.
\label{tab:splittingfunctions:two-loop}}
\end{table}

\end{appendix}

\bibliography{pTveto}

\providecommand{\href}[2]{#2}\begingroup\raggedright\begin{thebibliography}{10}

\bibitem{Banfi:2012yh}
A.~Banfi, G.P.~Salam and G.~Zanderighi, \emph{{NLL+NNLO predictions for
  jet-veto efficiencies in Higgs-boson and Drell-Yan production}},
  \href{https://doi.org/10.1007/JHEP06(2012)159}{\emph{JHEP} {\bfseries 06}
  (2012) 159} [\href{https://arxiv.org/abs/1203.5773}{{\ttfamily 1203.5773}}].

\bibitem{Becher:2012qa}
T.~Becher and M.~Neubert, \emph{{Factorization and NNLL Resummation for Higgs
  Production with a Jet Veto}},
  \href{https://doi.org/10.1007/JHEP07(2012)108}{\emph{JHEP} {\bfseries 07}
  (2012) 108} [\href{https://arxiv.org/abs/1205.3806}{{\ttfamily 1205.3806}}].

\bibitem{Tackmann:2012bt}
F.J.~Tackmann, J.R.~Walsh and S.~Zuberi, \emph{{Resummation Properties of Jet
  Vetoes at the LHC}},
  \href{https://doi.org/10.1103/PhysRevD.86.053011}{\emph{Phys. Rev. D}
  {\bfseries 86} (2012) 053011}
  [\href{https://arxiv.org/abs/1206.4312}{{\ttfamily 1206.4312}}].

\bibitem{Banfi:2012jm}
A.~Banfi, P.F.~Monni, G.P.~Salam and G.~Zanderighi, \emph{{Higgs and Z-boson
  production with a jet veto}},
  \href{https://doi.org/10.1103/PhysRevLett.109.202001}{\emph{Phys. Rev. Lett.}
  {\bfseries 109} (2012) 202001}
  [\href{https://arxiv.org/abs/1206.4998}{{\ttfamily 1206.4998}}].

\bibitem{Becher:2013xia}
T.~Becher, M.~Neubert and L.~Rothen, \emph{{Factorization and
  $N^{3}LL_{p}$+NNLO predictions for the Higgs cross section with a jet veto}},
  \href{https://doi.org/10.1007/JHEP10(2013)125}{\emph{JHEP} {\bfseries 10}
  (2013) 125} [\href{https://arxiv.org/abs/1307.0025}{{\ttfamily 1307.0025}}].

\bibitem{Stewart:2013faa}
I.W.~Stewart, F.J.~Tackmann, J.R.~Walsh and S.~Zuberi, \emph{{Jet $p_T$
  resummation in Higgs production at $NNLL'+NNLO$}},
  \href{https://doi.org/10.1103/PhysRevD.89.054001}{\emph{Phys. Rev. D}
  {\bfseries 89} (2014) 054001}
  [\href{https://arxiv.org/abs/1307.1808}{{\ttfamily 1307.1808}}].

\bibitem{Banfi:2015pju}
A.~Banfi, F.~Caola, F.A.~Dreyer, P.F.~Monni, G.P.~Salam, G.~Zanderighi et~al.,
  \emph{{Jet-vetoed Higgs cross section in gluon fusion at N$^{3}$LO+NNLL with
  small-$R$ resummation}},
  \href{https://doi.org/10.1007/JHEP04(2016)049}{\emph{JHEP} {\bfseries 04}
  (2016) 049} [\href{https://arxiv.org/abs/1511.02886}{{\ttfamily
  1511.02886}}].

\bibitem{Monni:2019yyr}
P.F.~Monni, L.~Rottoli and P.~Torrielli, \emph{{Higgs transverse momentum with
  a jet veto: a double-differential resummation}},
  \href{https://doi.org/10.1103/PhysRevLett.124.252001}{\emph{Phys. Rev. Lett.}
  {\bfseries 124} (2020) 252001}
  [\href{https://arxiv.org/abs/1909.04704}{{\ttfamily 1909.04704}}].

\bibitem{Becher:2014aya}
T.~Becher, R.~Frederix, M.~Neubert and L.~Rothen, \emph{{Automated NNLL $+$ NLO
  resummation for jet-veto cross sections}},
  \href{https://doi.org/10.1140/epjc/s10052-015-3368-y}{\emph{Eur. Phys. J. C}
  {\bfseries 75} (2015) 154} [\href{https://arxiv.org/abs/1412.8408}{{\ttfamily
  1412.8408}}].

\bibitem{Shao:2013uba}
D.Y.~Shao, C.S.~Li and H.T.~Li, \emph{{Resummation Prediction on Higgs and
  Vector Boson Associated Production with a Jet Veto at the LHC}},
  \href{https://doi.org/10.1007/JHEP02(2014)117}{\emph{JHEP} {\bfseries 02}
  (2014) 117} [\href{https://arxiv.org/abs/1309.5015}{{\ttfamily 1309.5015}}].

\bibitem{Li:2014ria}
Y.~Li and X.~Liu, \emph{{High precision predictions for exclusive $VH$
  production at the LHC}},
  \href{https://doi.org/10.1007/JHEP06(2014)028}{\emph{JHEP} {\bfseries 06}
  (2014) 028} [\href{https://arxiv.org/abs/1401.2149}{{\ttfamily 1401.2149}}].

\bibitem{Jaiswal:2014yba}
P.~Jaiswal and T.~Okui, \emph{{Explanation of the $WW$ excess at the LHC by
  jet-veto resummation}},
  \href{https://doi.org/10.1103/PhysRevD.90.073009}{\emph{Phys. Rev. D}
  {\bfseries 90} (2014) 073009}
  [\href{https://arxiv.org/abs/1407.4537}{{\ttfamily 1407.4537}}].

\bibitem{Wang:2015mvz}
Y.~Wang, C.S.~Li and Z.L.~Liu, \emph{{Resummation prediction on gauge boson
  pair production with a jet veto}},
  \href{https://doi.org/10.1103/PhysRevD.93.094020}{\emph{Phys. Rev. D}
  {\bfseries 93} (2016) 094020}
  [\href{https://arxiv.org/abs/1504.00509}{{\ttfamily 1504.00509}}].

\bibitem{Dawson:2016ysj}
S.~Dawson, P.~Jaiswal, Y.~Li, H.~Ramani and M.~Zeng, \emph{{Resummation of jet
  veto logarithms at N$^3$LL$_a$ + NNLO for $W^+ W^-$ production at the LHC}},
  \href{https://doi.org/10.1103/PhysRevD.94.114014}{\emph{Phys. Rev. D}
  {\bfseries 94} (2016) 114014}
  [\href{https://arxiv.org/abs/1606.01034}{{\ttfamily 1606.01034}}].

\bibitem{Tackmann:2016jyb}
F.J.~Tackmann, W.J.~Waalewijn and L.~Zeune, \emph{{Impact of Jet Veto
  Resummation on Slepton Searches}},
  \href{https://doi.org/10.1007/JHEP07(2016)119}{\emph{JHEP} {\bfseries 07}
  (2016) 119} [\href{https://arxiv.org/abs/1603.03052}{{\ttfamily
  1603.03052}}].

\bibitem{Ebert:2016idf}
M.A.~Ebert, S.~Liebler, I.~Moult, I.W.~Stewart, F.J.~Tackmann, K.~Tackmann
  et~al., \emph{{Exploiting jet binning to identify the initial state of
  high-mass resonances}},
  \href{https://doi.org/10.1103/PhysRevD.94.051901}{\emph{Phys. Rev. D}
  {\bfseries 94} (2016) 051901}
  [\href{https://arxiv.org/abs/1605.06114}{{\ttfamily 1605.06114}}].

\bibitem{Fuks:2017vtl}
B.~Fuks and R.~Ruiz, \emph{{A comprehensive framework for studying $W'$ and
  $Z'$ bosons at hadron colliders with automated jet veto resummation}},
  \href{https://doi.org/10.1007/JHEP05(2017)032}{\emph{JHEP} {\bfseries 05}
  (2017) 032} [\href{https://arxiv.org/abs/1701.05263}{{\ttfamily
  1701.05263}}].

\bibitem{Arpino:2019fmo}
L.~Arpino, A.~Banfi, S.~J\"ager and N.~Kauer, \emph{{BSM $WW$ production with a
  jet veto}}, \href{https://doi.org/10.1007/JHEP08(2019)076}{\emph{JHEP}
  {\bfseries 08} (2019) 076}
  [\href{https://arxiv.org/abs/1905.06646}{{\ttfamily 1905.06646}}].

\bibitem{Gangal:2014qda}
S.~Gangal, M.~Stahlhofen and F.J.~Tackmann, \emph{{Rapidity-Dependent Jet
  Vetoes}}, \href{https://doi.org/10.1103/PhysRevD.91.054023}{\emph{Phys. Rev.
  D} {\bfseries 91} (2015) 054023}
  [\href{https://arxiv.org/abs/1412.4792}{{\ttfamily 1412.4792}}].

\bibitem{Gangal:2020qik}
S.~Gangal, J.R.~Gaunt, F.J.~Tackmann and E.~Vryonidou, \emph{{Higgs Production
  at NNLL$'$+NNLO using Rapidity Dependent Jet Vetoes}},
  \href{https://doi.org/10.1007/JHEP05(2020)054}{\emph{JHEP} {\bfseries 05}
  (2020) 054} [\href{https://arxiv.org/abs/2003.04323}{{\ttfamily
  2003.04323}}].

\bibitem{Hornig:2017pud}
A.~Hornig, D.~Kang, Y.~Makris and T.~Mehen, \emph{{Transverse Vetoes with
  Rapidity Cutoff in SCET}},
  \href{https://doi.org/10.1007/JHEP12(2017)043}{\emph{JHEP} {\bfseries 12}
  (2017) 043} [\href{https://arxiv.org/abs/1708.08467}{{\ttfamily
  1708.08467}}].

\bibitem{Michel:2018hui}
J.K.L.~Michel, P.~Pietrulewicz and F.J.~Tackmann, \emph{{Jet Veto Resummation
  with Jet Rapidity Cuts}},
  \href{https://doi.org/10.1007/JHEP04(2019)142}{\emph{JHEP} {\bfseries 04}
  (2019) 142} [\href{https://arxiv.org/abs/1810.12911}{{\ttfamily
  1810.12911}}].

\bibitem{Zeng:2015iba}
M.~Zeng, \emph{{Drell-Yan process with jet vetoes: breaking of generalized
  factorization}}, \href{https://doi.org/10.1007/JHEP10(2015)189}{\emph{JHEP}
  {\bfseries 10} (2015) 189}
  [\href{https://arxiv.org/abs/1507.01652}{{\ttfamily 1507.01652}}].

\bibitem{Bauer:2000yr}
C.W.~Bauer, S.~Fleming, D.~Pirjol and I.W.~Stewart, \emph{{An Effective field
  theory for collinear and soft gluons: Heavy to light decays}},
  \href{https://doi.org/10.1103/PhysRevD.63.114020}{\emph{Phys. Rev. D}
  {\bfseries 63} (2001) 114020}
  [\href{https://arxiv.org/abs/hep-ph/0011336}{{\ttfamily hep-ph/0011336}}].

\bibitem{Bauer:2001yt}
C.W.~Bauer, D.~Pirjol and I.W.~Stewart, \emph{{Soft collinear factorization in
  effective field theory}},
  \href{https://doi.org/10.1103/PhysRevD.65.054022}{\emph{Phys. Rev. D}
  {\bfseries 65} (2002) 054022}
  [\href{https://arxiv.org/abs/hep-ph/0109045}{{\ttfamily hep-ph/0109045}}].

\bibitem{Beneke:2002ph}
M.~Beneke, A.P.~Chapovsky, M.~Diehl and T.~Feldmann, \emph{{Soft collinear
  effective theory and heavy to light currents beyond leading power}},
  \href{https://doi.org/10.1016/S0550-3213(02)00687-9}{\emph{Nucl. Phys. B}
  {\bfseries 643} (2002) 431}
  [\href{https://arxiv.org/abs/hep-ph/0206152}{{\ttfamily hep-ph/0206152}}].

\bibitem{Becher:2010tm}
T.~Becher and M.~Neubert, \emph{{Drell-Yan Production at Small $q_T$,
  Transverse Parton Distributions and the Collinear Anomaly}},
  \href{https://doi.org/10.1140/epjc/s10052-011-1665-7}{\emph{Eur. Phys. J. C}
  {\bfseries 71} (2011) 1665}
  [\href{https://arxiv.org/abs/1007.4005}{{\ttfamily 1007.4005}}].

\bibitem{Becher:2012yn}
T.~Becher, M.~Neubert and D.~Wilhelm, \emph{{Higgs-Boson Production at Small
  Transverse Momentum}},
  \href{https://doi.org/10.1007/JHEP05(2013)110}{\emph{JHEP} {\bfseries 05}
  (2013) 110} [\href{https://arxiv.org/abs/1212.2621}{{\ttfamily 1212.2621}}].

\bibitem{Catani:2013tia}
S.~Catani, L.~Cieri, D.~de~Florian, G.~Ferrera and M.~Grazzini,
  \emph{{Universality of transverse-momentum resummation and hard factors at
  the NNLO}},
  \href{https://doi.org/10.1016/j.nuclphysb.2014.02.011}{\emph{Nucl. Phys. B}
  {\bfseries 881} (2014) 414}
  [\href{https://arxiv.org/abs/1311.1654}{{\ttfamily 1311.1654}}].

\bibitem{Gehrmann:2014yya}
T.~Gehrmann, T.~Luebbert and L.L.~Yang, \emph{{Calculation of the transverse
  parton distribution functions at next-to-next-to-leading order}},
  \href{https://doi.org/10.1007/JHEP06(2014)155}{\emph{JHEP} {\bfseries 06}
  (2014) 155} [\href{https://arxiv.org/abs/1403.6451}{{\ttfamily 1403.6451}}].

\bibitem{Lubbert:2016rku}
T.~Luebbert, J.~Oredsson and M.~Stahlhofen, \emph{{Rapidity renormalized TMD
  soft and beam functions at two loops}},
  \href{https://doi.org/10.1007/JHEP03(2016)168}{\emph{JHEP} {\bfseries 03}
  (2016) 168} [\href{https://arxiv.org/abs/1602.01829}{{\ttfamily
  1602.01829}}].

\bibitem{Echevarria:2016scs}
M.G.~Echevarria, I.~Scimemi and A.~Vladimirov, \emph{{Unpolarized Transverse
  Momentum Dependent Parton Distribution and Fragmentation Functions at
  next-to-next-to-leading order}},
  \href{https://doi.org/10.1007/JHEP09(2016)004}{\emph{JHEP} {\bfseries 09}
  (2016) 004} [\href{https://arxiv.org/abs/1604.07869}{{\ttfamily
  1604.07869}}].

\bibitem{Luo:2019hmp}
M.-X.~Luo, X.~Wang, X.~Xu, L.L.~Yang, T.-Z.~Yang and H.X.~Zhu,
  \emph{{Transverse Parton Distribution and Fragmentation Functions at NNLO:
  the Quark Case}}, \href{https://doi.org/10.1007/JHEP10(2019)083}{\emph{JHEP}
  {\bfseries 10} (2019) 083}
  [\href{https://arxiv.org/abs/1908.03831}{{\ttfamily 1908.03831}}].

\bibitem{Luo:2019bmw}
M.-X.~Luo, T.-Z.~Yang, H.X.~Zhu and Y.J.~Zhu, \emph{{Transverse Parton
  Distribution and Fragmentation Functions at NNLO: the Gluon Case}},
  \href{https://doi.org/10.1007/JHEP01(2020)040}{\emph{JHEP} {\bfseries 01}
  (2020) 040} [\href{https://arxiv.org/abs/1909.13820}{{\ttfamily
  1909.13820}}].

\bibitem{Luo:2019szz}
M.-x.~Luo, T.-Z.~Yang, H.X.~Zhu and Y.J.~Zhu, \emph{{Quark Transverse Parton
  Distribution at the Next-to-Next-to-Next-to-Leading Order}},
  \href{https://doi.org/10.1103/PhysRevLett.124.092001}{\emph{Phys. Rev. Lett.}
  {\bfseries 124} (2020) 092001}
  [\href{https://arxiv.org/abs/1912.05778}{{\ttfamily 1912.05778}}].

\bibitem{Ebert:2020yqt}
M.A.~Ebert, B.~Mistlberger and G.~Vita, \emph{{Transverse momentum dependent
  PDFs at N$^3$LO}}, \href{https://doi.org/10.1007/JHEP09(2020)146}{\emph{JHEP}
  {\bfseries 09} (2020) 146}
  [\href{https://arxiv.org/abs/2006.05329}{{\ttfamily 2006.05329}}].

\bibitem{Luo:2020epw}
M.-x.~Luo, T.-Z.~Yang, H.X.~Zhu and Y.J.~Zhu, \emph{{Unpolarized quark and
  gluon TMD PDFs and FFs at N$^{3}$LO}},
  \href{https://doi.org/10.1007/JHEP06(2021)115}{\emph{JHEP} {\bfseries 06}
  (2021) 115} [\href{https://arxiv.org/abs/2012.03256}{{\ttfamily
  2012.03256}}].

\bibitem{Stewart:2010qs}
I.W.~Stewart, F.J.~Tackmann and W.J.~Waalewijn, \emph{{The Quark Beam Function
  at NNLL}}, \href{https://doi.org/10.1007/JHEP09(2010)005}{\emph{JHEP}
  {\bfseries 09} (2010) 005} [\href{https://arxiv.org/abs/1002.2213}{{\ttfamily
  1002.2213}}].

\bibitem{Berger:2010xi}
C.F.~Berger, C.~Marcantonini, I.W.~Stewart, F.J.~Tackmann and W.J.~Waalewijn,
  \emph{{Higgs Production with a Central Jet Veto at NNLL+NNLO}},
  \href{https://doi.org/10.1007/JHEP04(2011)092}{\emph{JHEP} {\bfseries 04}
  (2011) 092} [\href{https://arxiv.org/abs/1012.4480}{{\ttfamily 1012.4480}}].

\bibitem{Gaunt:2014xga}
J.R.~Gaunt, M.~Stahlhofen and F.J.~Tackmann, \emph{{The Quark Beam Function at
  Two Loops}}, \href{https://doi.org/10.1007/JHEP04(2014)113}{\emph{JHEP}
  {\bfseries 04} (2014) 113} [\href{https://arxiv.org/abs/1401.5478}{{\ttfamily
  1401.5478}}].

\bibitem{Gaunt:2014cfa}
J.~Gaunt, M.~Stahlhofen and F.J.~Tackmann, \emph{{The Gluon Beam Function at
  Two Loops}}, \href{https://doi.org/10.1007/JHEP08(2014)020}{\emph{JHEP}
  {\bfseries 08} (2014) 020} [\href{https://arxiv.org/abs/1405.1044}{{\ttfamily
  1405.1044}}].

\bibitem{Melnikov:2018jxb}
K.~Melnikov, R.~Rietkerk, L.~Tancredi and C.~Wever, \emph{{Double-real
  contribution to the quark beam function at N$^{3}$LO QCD}},
  \href{https://doi.org/10.1007/JHEP02(2019)159}{\emph{JHEP} {\bfseries 02}
  (2019) 159} [\href{https://arxiv.org/abs/1809.06300}{{\ttfamily
  1809.06300}}].

\bibitem{Melnikov:2019pdm}
K.~Melnikov, R.~Rietkerk, L.~Tancredi and C.~Wever, \emph{{Triple-real
  contribution to the quark beam function in QCD at
  next-to-next-to-next-to-leading order}},
  \href{https://doi.org/10.1007/JHEP06(2019)033}{\emph{JHEP} {\bfseries 06}
  (2019) 033} [\href{https://arxiv.org/abs/1904.02433}{{\ttfamily
  1904.02433}}].

\bibitem{Behring:2019quf}
A.~Behring, K.~Melnikov, R.~Rietkerk, L.~Tancredi and C.~Wever, \emph{{Quark
  beam function at next-to-next-to-next-to-leading order in perturbative QCD in
  the generalized large-$N_c$ approximation}},
  \href{https://doi.org/10.1103/PhysRevD.100.114034}{\emph{Phys. Rev. D}
  {\bfseries 100} (2019) 114034}
  [\href{https://arxiv.org/abs/1910.10059}{{\ttfamily 1910.10059}}].

\bibitem{Baranowski:2020xlp}
D.~Baranowski, \emph{{NNLO zero-jettiness beam and soft functions to higher
  orders in the dimensional-regularization parameter $\epsilon$}},
  \href{https://doi.org/10.1140/epjc/s10052-020-8047-y}{\emph{Eur. Phys. J. C}
  {\bfseries 80} (2020) 523}
  [\href{https://arxiv.org/abs/2004.03285}{{\ttfamily 2004.03285}}].

\bibitem{Ebert:2020unb}
M.A.~Ebert, B.~Mistlberger and G.~Vita, \emph{{$N$-jettiness beam functions at
  N$^{3}$LO}}, \href{https://doi.org/10.1007/JHEP09(2020)143}{\emph{JHEP}
  {\bfseries 09} (2020) 143}
  [\href{https://arxiv.org/abs/2006.03056}{{\ttfamily 2006.03056}}].

\bibitem{Jain:2011iu}
A.~Jain, M.~Procura and W.J.~Waalewijn, \emph{{Fully-Unintegrated Parton
  Distribution and Fragmentation Functions at Perturbative $k_T$}},
  \href{https://doi.org/10.1007/JHEP04(2012)132}{\emph{JHEP} {\bfseries 04}
  (2012) 132} [\href{https://arxiv.org/abs/1110.0839}{{\ttfamily 1110.0839}}].

\bibitem{Gaunt:2014xxa}
J.R.~Gaunt and M.~Stahlhofen, \emph{{The Fully-Differential Quark Beam Function
  at NNLO}}, \href{https://doi.org/10.1007/JHEP12(2014)146}{\emph{JHEP}
  {\bfseries 12} (2014) 146} [\href{https://arxiv.org/abs/1409.8281}{{\ttfamily
  1409.8281}}].

\bibitem{Gaunt:2020xlc}
J.R.~Gaunt and M.~Stahlhofen, \emph{{The fully-differential gluon beam function
  at NNLO}}, \href{https://doi.org/10.1007/JHEP07(2020)234}{\emph{JHEP}
  {\bfseries 07} (2020) 234}
  [\href{https://arxiv.org/abs/2004.11915}{{\ttfamily 2004.11915}}].

\bibitem{Gangal:2016kuo}
S.~Gangal, J.R.~Gaunt, M.~Stahlhofen and F.J.~Tackmann, \emph{{Two-Loop Beam
  and Soft Functions for Rapidity-Dependent Jet Vetoes}},
  \href{https://doi.org/10.1007/JHEP02(2017)026}{\emph{JHEP} {\bfseries 02}
  (2017) 026} [\href{https://arxiv.org/abs/1608.01999}{{\ttfamily
  1608.01999}}].

\bibitem{Bell:2018gce}
G.~Bell, A.~Hornig, C.~Lee and J.~Talbert, \emph{{$e^+ e^-$ angularity
  distributions at NNLL$^\prime$ accuracy}},
  \href{https://doi.org/10.1007/JHEP01(2019)147}{\emph{JHEP} {\bfseries 01}
  (2019) 147} [\href{https://arxiv.org/abs/1808.07867}{{\ttfamily
  1808.07867}}].

\bibitem{Becher:2011dz}
T.~Becher and G.~Bell, \emph{{Analytic Regularization in Soft-Collinear
  Effective Theory}},
  \href{https://doi.org/10.1016/j.physletb.2012.05.016}{\emph{Phys. Lett. B}
  {\bfseries 713} (2012) 41} [\href{https://arxiv.org/abs/1112.3907}{{\ttfamily
  1112.3907}}].

\bibitem{Bell:2021dpb}
G.~Bell, K.~Brune, G.~Das and M.~Wald, \emph{{Automation of Beam and Jet
  functions at NNLO}},  in \emph{{15th International Symposium on Radiative
  Corrections: Applications of Quantum Field Theory to Phenomenology AND
  LoopFest XIX: Workshop on Radiative Corrections for the LHC and Future
  Colliders}}, 10, 2021 [\href{https://arxiv.org/abs/2110.04804}{{\ttfamily
  2110.04804}}].

\bibitem{BBDW2}
G.~Bell, K.~Brune, G.~Das and M.~Wald, \emph{{Automated calculation of Beam
  functions at NNLO}},  in \emph{{Loops and Legs in Quantum Field Theory -
  LL2022, 25-30 April 2022, Ettal, Germany}}.

\bibitem{Bell:2018vaa}
G.~Bell, R.~Rahn and J.~Talbert, \emph{{Two-loop anomalous dimensions of
  generic dijet soft functions}},
  \href{https://doi.org/10.1016/j.nuclphysb.2018.09.026}{\emph{Nucl. Phys. B}
  {\bfseries 936} (2018) 520}
  [\href{https://arxiv.org/abs/1805.12414}{{\ttfamily 1805.12414}}].

\bibitem{Bell:2018oqa}
G.~Bell, R.~Rahn and J.~Talbert, \emph{{Generic dijet soft functions at
  two-loop order: correlated emissions}},
  \href{https://doi.org/10.1007/JHEP07(2019)101}{\emph{JHEP} {\bfseries 07}
  (2019) 101} [\href{https://arxiv.org/abs/1812.08690}{{\ttfamily
  1812.08690}}].

\bibitem{Bell:2020yzz}
G.~Bell, R.~Rahn and J.~Talbert, \emph{{Generic dijet soft functions at
  two-loop order: uncorrelated emissions}},
  \href{https://doi.org/10.1007/JHEP09(2020)015}{\emph{JHEP} {\bfseries 09}
  (2020) 015} [\href{https://arxiv.org/abs/2004.08396}{{\ttfamily
  2004.08396}}].

\bibitem{Bell:2018mkk}
G.~Bell, B.~Dehnadi, T.~Mohrmann and R.~Rahn, \emph{{Automated Calculation of
  ${\pmb N}$-jet Soft Functions}},
  \href{https://doi.org/10.22323/1.303.0044}{\emph{PoS} {\bfseries LL2018}
  (2018) 044} [\href{https://arxiv.org/abs/1808.07427}{{\ttfamily
  1808.07427}}].

\bibitem{Becher:2011pf}
T.~Becher, G.~Bell and M.~Neubert, \emph{{Factorization and Resummation for Jet
  Broadening}},
  \href{https://doi.org/10.1016/j.physletb.2011.09.005}{\emph{Phys. Lett. B}
  {\bfseries 704} (2011) 276}
  [\href{https://arxiv.org/abs/1104.4108}{{\ttfamily 1104.4108}}].

\bibitem{Ritzmann:2014mka}
M.~Ritzmann and W.J.~Waalewijn, \emph{{Fragmentation in Jets at NNLO}},
  \href{https://doi.org/10.1103/PhysRevD.90.054029}{\emph{Phys. Rev. D}
  {\bfseries 90} (2014) 054029}
  [\href{https://arxiv.org/abs/1407.3272}{{\ttfamily 1407.3272}}].

\bibitem{Kosower:1999rx}
D.A.~Kosower and P.~Uwer, \emph{{One loop splitting amplitudes in gauge
  theory}}, \href{https://doi.org/10.1016/S0550-3213(99)00583-0}{\emph{Nucl.
  Phys. B} {\bfseries 563} (1999) 477}
  [\href{https://arxiv.org/abs/hep-ph/9903515}{{\ttfamily hep-ph/9903515}}].

\bibitem{Bern:1999ry}
Z.~Bern, V.~Del~Duca, W.B.~Kilgore and C.R.~Schmidt, \emph{{The infrared
  behavior of one loop QCD amplitudes at next-to-next-to leading order}},
  \href{https://doi.org/10.1103/PhysRevD.60.116001}{\emph{Phys. Rev. D}
  {\bfseries 60} (1999) 116001}
  [\href{https://arxiv.org/abs/hep-ph/9903516}{{\ttfamily hep-ph/9903516}}].

\bibitem{Sborlini:2013jba}
G.F.R.~Sborlini, D.~de~Florian and G.~Rodrigo, \emph{{Double collinear
  splitting amplitudes at next-to-leading order}},
  \href{https://doi.org/10.1007/JHEP01(2014)018}{\emph{JHEP} {\bfseries 01}
  (2014) 018} [\href{https://arxiv.org/abs/1310.6841}{{\ttfamily 1310.6841}}].

\bibitem{Campbell:1997hg}
J.M.~Campbell and E.W.N.~Glover, \emph{{Double unresolved approximations to
  multiparton scattering amplitudes}},
  \href{https://doi.org/10.1016/S0550-3213(98)00295-8}{\emph{Nucl. Phys. B}
  {\bfseries 527} (1998) 264}
  [\href{https://arxiv.org/abs/hep-ph/9710255}{{\ttfamily hep-ph/9710255}}].

\bibitem{Catani:1998nv}
S.~Catani and M.~Grazzini, \emph{{Collinear factorization and splitting
  functions for next-to-next-to-leading order QCD calculations}},
  \href{https://doi.org/10.1016/S0370-2693(98)01513-5}{\emph{Phys. Lett. B}
  {\bfseries 446} (1999) 143}
  [\href{https://arxiv.org/abs/hep-ph/9810389}{{\ttfamily hep-ph/9810389}}].

\bibitem{Borowka:2017idc}
S.~Borowka, G.~Heinrich, S.~Jahn, S.P.~Jones, M.~Kerner, J.~Schlenk et~al.,
  \emph{{pySecDec: a toolbox for the numerical evaluation of multi-scale
  integrals}}, \href{https://doi.org/10.1016/j.cpc.2017.09.015}{\emph{Comput.
  Phys. Commun.} {\bfseries 222} (2018) 313}
  [\href{https://arxiv.org/abs/1703.09692}{{\ttfamily 1703.09692}}].

\bibitem{Hahn:2004fe}
T.~Hahn, \emph{{CUBA: A Library for multidimensional numerical integration}},
  \href{https://doi.org/10.1016/j.cpc.2005.01.010}{\emph{Comput. Phys. Commun.}
  {\bfseries 168} (2005) 78}
  [\href{https://arxiv.org/abs/hep-ph/0404043}{{\ttfamily hep-ph/0404043}}].

\bibitem{BBDW}
G.~Bell, K.~Brune, G.~Das and M.~Wald, \emph{{Generic beam functions at
  two-loop order}}, {\emph{\normalfont{in preparation}} }.

\bibitem{Abreu:2022zgo}
S.~Abreu, J.R.~Gaunt, P.F.~Monni, L.~Rottoli and R.~Szafron, \emph{{Quark and
  gluon two-loop beam functions for leading-jet $p_T$ and slicing at NNLO}},
  \href{https://arxiv.org/abs/2207.07037}{{\ttfamily 2207.07037}}.

\bibitem{Becher:2006mr}
T.~Becher, M.~Neubert and B.D.~Pecjak, \emph{{Factorization and Momentum-Space
  Resummation in Deep-Inelastic Scattering}},
  \href{https://doi.org/10.1088/1126-6708/2007/01/076}{\emph{JHEP} {\bfseries
  01} (2007) 076} [\href{https://arxiv.org/abs/hep-ph/0607228}{{\ttfamily
  hep-ph/0607228}}].

\bibitem{Moch:2004pa}
S.~Moch, J.A.M.~Vermaseren and A.~Vogt, \emph{{The Three loop splitting
  functions in QCD: The Nonsinglet case}},
  \href{https://doi.org/10.1016/j.nuclphysb.2004.03.030}{\emph{Nucl. Phys. B}
  {\bfseries 688} (2004) 101}
  [\href{https://arxiv.org/abs/hep-ph/0403192}{{\ttfamily hep-ph/0403192}}].

\bibitem{Vogt:2004mw}
A.~Vogt, S.~Moch and J.A.M.~Vermaseren, \emph{{The Three-loop splitting
  functions in QCD: The Singlet case}},
  \href{https://doi.org/10.1016/j.nuclphysb.2004.04.024}{\emph{Nucl. Phys. B}
  {\bfseries 691} (2004) 129}
  [\href{https://arxiv.org/abs/hep-ph/0404111}{{\ttfamily hep-ph/0404111}}].

\bibitem{Blumlein:2021enk}
J.~Bl\"umlein, P.~Marquard, C.~Schneider and K.~Sch\"onwald, \emph{{The
  three-loop unpolarized and polarized non-singlet anomalous dimensions from
  off shell operator matrix elements}},
  \href{https://doi.org/10.1016/j.nuclphysb.2021.115542}{\emph{Nucl. Phys. B}
  {\bfseries 971} (2021) 115542}
  [\href{https://arxiv.org/abs/2107.06267}{{\ttfamily 2107.06267}}].

\bibitem{Curci:1980uw}
G.~Curci, W.~Furmanski and R.~Petronzio, \emph{{Evolution of Parton Densities
  Beyond Leading Order: The Nonsinglet Case}},
  \href{https://doi.org/10.1016/0550-3213(80)90003-6}{\emph{Nucl. Phys. B}
  {\bfseries 175} (1980) 27}.

\bibitem{Furmanski:1980cm}
W.~Furmanski and R.~Petronzio, \emph{{Singlet Parton Densities Beyond Leading
  Order}}, \href{https://doi.org/10.1016/0370-2693(80)90636-X}{\emph{Phys.
  Lett. B} {\bfseries 97} (1980) 437}.

\bibitem{Ellis:1996nn}
R.K.~Ellis and W.~Vogelsang, \emph{{The Evolution of parton distributions
  beyond leading order: The Singlet case}},
  \href{https://arxiv.org/abs/hep-ph/9602356}{{\ttfamily hep-ph/9602356}}.

\end{thebibliography}\endgroup

\end{document}